\documentclass{PoS}
\usepackage{wrapfig}
\usepackage{multicol}
\usepackage{lineno}
\usepackage{amsmath}

\title{Muon Calibration at SoLid}

\ShortTitle{Muon Calibration at SoLid}

\author{\speaker{Daniel Saunders}\thanks{On behalf of the SoLid collaboration.}\\
        University of Bristol\\
        E-mail: \email{dan.saunders@bristol.ac.uk}}

\abstract{The SoLid experiment aims to make a measurement of very short distance neutrino oscillations using reactor antineutrinos. Key to its sensitivity are the experiment's high spatial and energy resolution, combined with a very suitable reactor source and efficient background rejection. The fine segmentation of the detector (cubes of side 5cm), and ability to resolve signals in space and time, gives SoLid the capability to track cosmic muons. In principle a source of background, these turn into a valuable calibration source if they can be cleanly identified. This work presents the first energy calibration results, using cosmic muons, of the 288kg SoLid prototype SM1. This includes the methodology of tracking at SoLid, cosmic ray angular analyses at the reactor site, estimates of the time resolution, and calibrations at the cube level.}

\FullConference{The European Physical Society Conference on High Energy Physics\\
		22--29 July 2015\\
		Vienna, Austria}

\begin{document}
\section{Introduction}
The SoLid experiment aims to investigate the reactor anomaly. Previous experiments have shown a deficit in the neutrino flux from nuclear reactors to the level of $2.5\sigma$ \cite{gMention}. Possible explanations include the existence of a sterile neutrino at the mass scale $\Delta m_s^2 \approx 1$ eV$^2$. The SoLid experiment will search for this neutrino starting in the second half of 2016, and will run until 2020. The results outlined below concern the most recent SoLid prototype, SM1, which has $10\%$ (288 kg) of the final planned mass. SM1 took data in 2015 under several running conditions, including: reactor-on, reactor-off, as well as radioactive source calibrations with both photons and neutrons. 

Of key importance to this search is a detector capable of efficient background rejection and high energy resolution. The SoLid detector has been designed with both these aspects in mind, and a more detailed outline of the detector design is given in \cite{celinePre} and \cite{nickPre}. SM1 is composed of Polyvinyl Toluene (PVT) scintillating cubes of 5 cm side length, with grooves for light fibres, placed in an arrangement of 9 planes. One face of each cube is lined with a sheet of $^6$LiF:ZnS, allowing for the detection of neutrons via scintillation light \cite{simonPre}. Each plane contains $16 \times 16$ cubes (see Figure~\ref{fig:muon_event_display}). For a plane, each row and column of cubes is coupled to a wavelength shifting light fibre, providing vertical and horizontal read out. A silicon photomultiplier is attached to one of the ends of each fibre, with a mirror is placed at the other end, giving a total of 288 channels. The segmentation of the detector allows for a detailed understanding of backgrounds and event topologies. Neutrinos interact with the detector via inverse beta decay (IBD) - the neutrino interacts with a proton, producing a positron signal and a delayed neutron signal (mean delay of $50 \mu s$). Thus, a good understanding of the timing performance of the detector is also required. A detailed account of the neutron capture mechanism, signal and identification is outlined in \cite{simonPre}.  

A significant source of background is from cosmic muons that can produce fast neutrons, and given the additional gamma radiation background from the reactor, this can lead to false IBD-like events. However, the fine segmentation of the detector allows most of these cosmic events to be reconstructed. The muon energy deposition distribution provides a standard candle that can be used for channel and cube equalisation, and by comparison to simulation, the absolute energy scale can be extracted. Further, since the time taken for a muon to pass through the detector is well known, these events can also be used to monitor the timing stability of the detector.

This article is ordered as follows: Section~\ref{sec:muon_tracking} outlines muon reconstruction at SoLid. Section~\ref{sec:energy_calib} outlines the energy calibration procedures developed for the detector, including channel and cube corrections. Finally, in Section~\ref{sec:timing_resolution}, the timing resolution of the detector will be presented. Reactor-off data is used throughout. 

\section{Muon Tracking}
\label{sec:muon_tracking}
Tracks at SoLid at formed from track hits, which are defined as signals from light fibres induced by tracks. Examples of these signals are shown in ~\cite{nickPre}. In all of the following, only tracks with 8 or more track hits are considered, allowing for their paths through the detector to be accurately reconstructed. These are known as \emph{long tracks}. An example from data is shown in the event display shown in Figure~\ref{fig:muon_event_display}. Tracks formed from fewer tracks hits are also reconstructed (short tracks), and are used in other analyses (e.g. as a muon veto).

Long tracks are identified by looking for large numbers of track hits in co-incidence within a time window from both sets of fibres (horizontal and vertical). The value of this time window $\Delta t _{Tracking}$ is set by the timing resolution of the detector for electromagnetic signals - see Section~\ref{sec:timing_resolution}. The default tracking cuts are:

\begin{figure}
\begin{center}
\includegraphics[width=0.75\columnwidth]{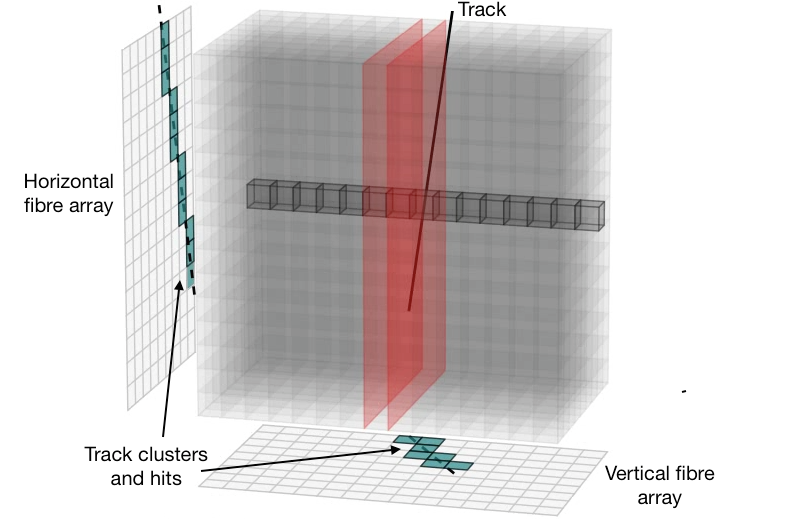}
\caption{SoLid prototype SM1. Displayed is a long track event from reactor-off data. The track hits are shown as projections. In this event, whilst intercepting the marked horizontal row of cubes, the track is contained within one cube (confining planes highlighted in red). These cases are used for energy calibrations.}
\label{fig:muon_event_display}
\end{center}
\end{figure}

\begin{center}
$\Delta t_{Tracking} = 50$ ns, and $n_{Track}$ $_{Hits} \geqslant 7 $
\end{center}

\begin{figure}
\begin{center}
\includegraphics[width=0.8\columnwidth]{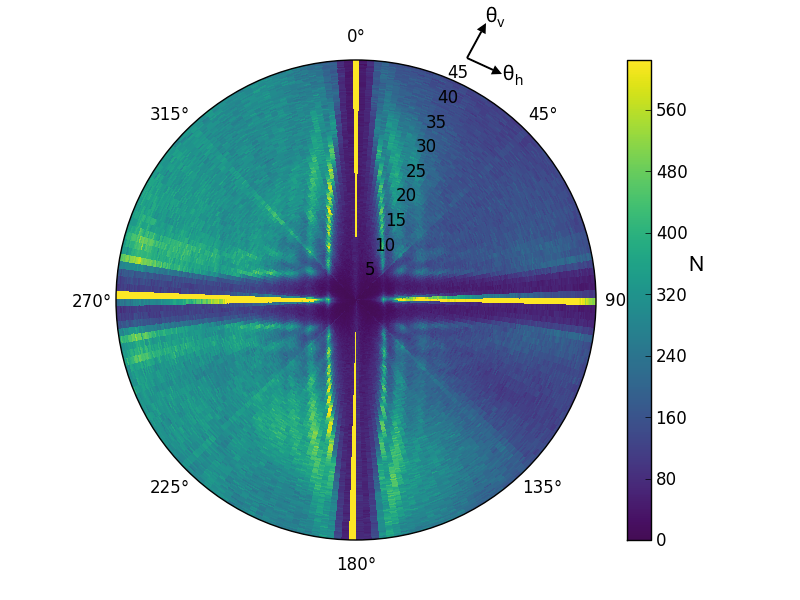}
\caption{Long track angular distribution in SM1. The cross-shaped yellow lines at right angles are due to the segmentation of the detector, where the angular resolution is decreased due to tracks passing through a single plane of cubes. The reactor is towards the east ($90^{\circ}$), providing extra shielding that is observed. Bins have equal solid angle size.}
\label{fig:angular_dist}
\end{center}
\end{figure}

Further cuts on track parameters are used later to select tracks with high quality reconstruction. The angular distribution of long tracks in SM1 is shown in Figure~\ref{fig:angular_dist}. The effects of segmentation are clearly observed, as well as expected scattering patterns given the material near the detector, including shielding due to the reactor. 

The geometric acceptance of this method is estimated using simulation. Tracks are generated with segmentation effects applied (due to the detector geometry) to investigate the effect of segmentation on resolution only - other effects such as detector response are not be considered. Using the above tracking cuts, the fraction of simulated tracks tagged as a long track is in the range: [0.75, 0.4], for the track steepness interval (measured from the vertical): $[0^{\circ},$ $45^{\circ}]$. 

\section{Energy Calibration}
\label{sec:energy_calib}
The main factors contributing to variations in energy calibration between cubes in SoLid are:
\begin{itemize}
\item Channel-to-Channel variations (e.g. due to gain differences in the electronic read out).
\item Attenuation loss in the fibre.
\item Cube-to-Cube variations (e.g. in the coupling between the fibres and cubes).
\end{itemize}

 The quantity used in each calibration step is the energy deposited per unit path length $dE/dx$. This quantity has the advantage of being well described by simulation, providing means to extract the energy scale of the detector. 
 
 In data, $dE/dx$ is found by considering muon events such as that in Figure~\ref{fig:muon_event_display}. $dE \approx \Delta E$ is taken as the integral of the signal deposited in the fibre, and $dx \approx \Delta x$ is taken as the path-length through the fibre's row/column of cubes. Only tracks where the reconstructed muon passes through exactly one cube in the given row/column are used. This condition is true for $40\%$ of track hits, and allows a $dE/dx$ distribution to be found for each cube. Simulation (using Geant4 \cite{geant4}) predicts the $dE/dx$ distribution to be closely approximated by a Landau distribution, whose most probable value $MPV_{Sim} = 1.78$ MeV/cm. To extract the energy scale, signal integrals are linearly scaled such that the $MPV_{Data} = MPV_{Sim}$. 
 
\begin{wrapfigure}{R}{0.5\textwidth}
\includegraphics[width=0.5\textwidth]{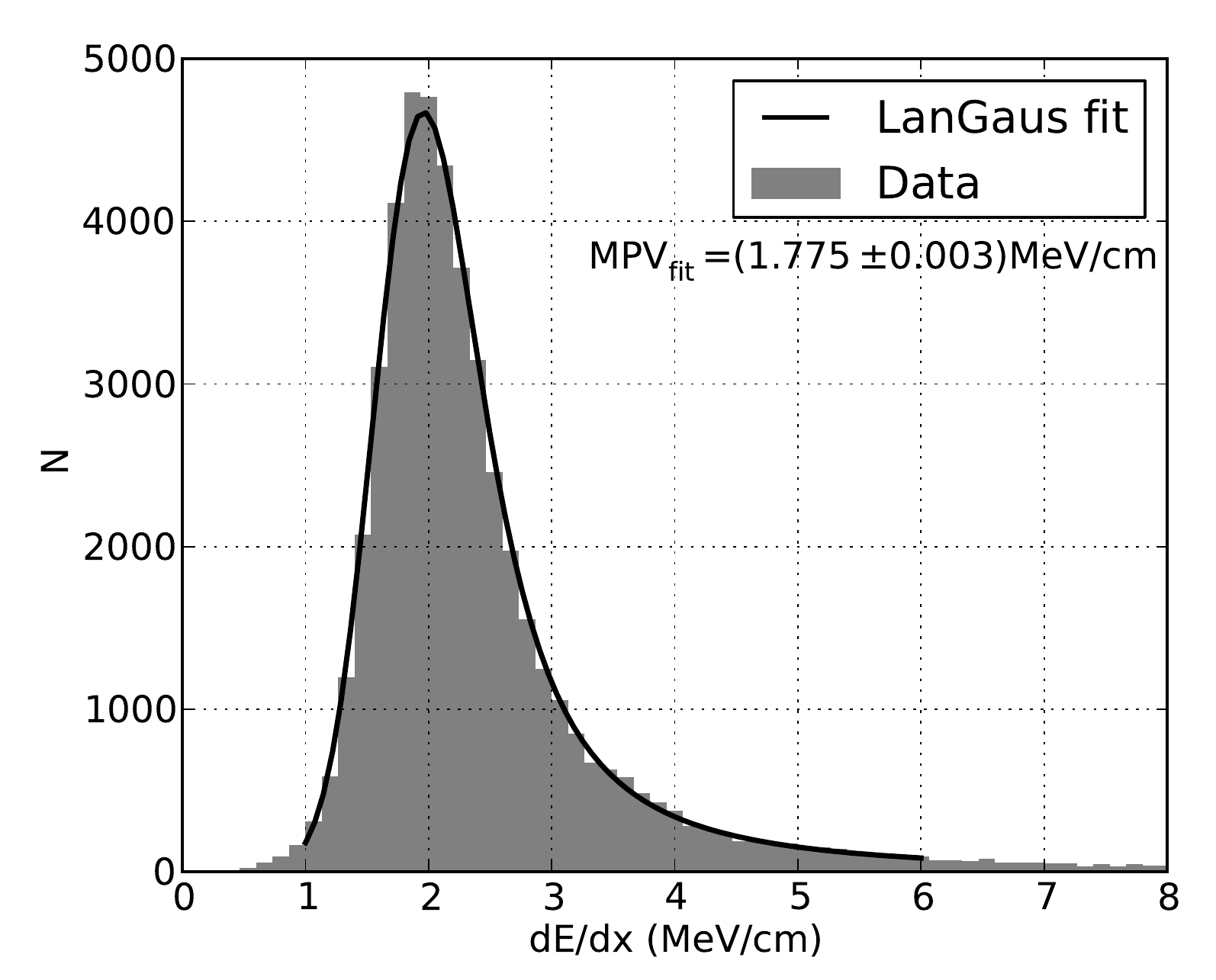}
\caption{$dE/dx$ distribution of tracks passing through all cubes at SM1. The data is fitted with a Landau $\otimes$ Gaussian convolution.}
\label{fig:dEdx_sm1}
\end{wrapfigure}

\subsection{Path Length Resolution Optimisation}
Further selection criteria are used to ensure tracks used for calibration are of good quality. In particular, comparisons with simulation show that tracks angled from the vertical and horizontal (i.e. diagonal tracks) are reconstructed with better accuracy. Specifically, cuts are placed on the minimum size of the clusters formed from vertical and horizontal track hits, and tracks are selected if both clusters have size $\geqslant 7$. Furthermore, an tuned track fit algorithm is used. This is a weighted linear regression, where the weight for each track hit is proportional to the signal integral of the track hit, raised to some power $\alpha$. Comparison between simulation and real data gives an optimum value of $\alpha = 1.6 \pm 0.1$ (stat). With these optimisations, the path length reconstruction resolution for tracks passing through cubes is $\sigma_{Path}$ $_{Length} = 1.2$ mm.

An example $dE/dx$ distribution using all cubes in SM1 is shown in Figure~\ref{fig:dEdx_sm1}. Each calibration step is now outlined in turn.

\subsection{Channel-to-Channel Calibration}
The method outlined above for finding $dE/dx$ can be performed for each channel independently by integrating over all the cubes read out by that channel (thus, this is similar to a common mode correction). A gain calibration constant is introduced for each channel that linearly scales the signals on that channel. The calibration constant of each channel is floated to scale its $dE/dx$ distribution to an arbitrary reference channel.

\begin{figure}
\begin{center}
\includegraphics[width=0.9\columnwidth]{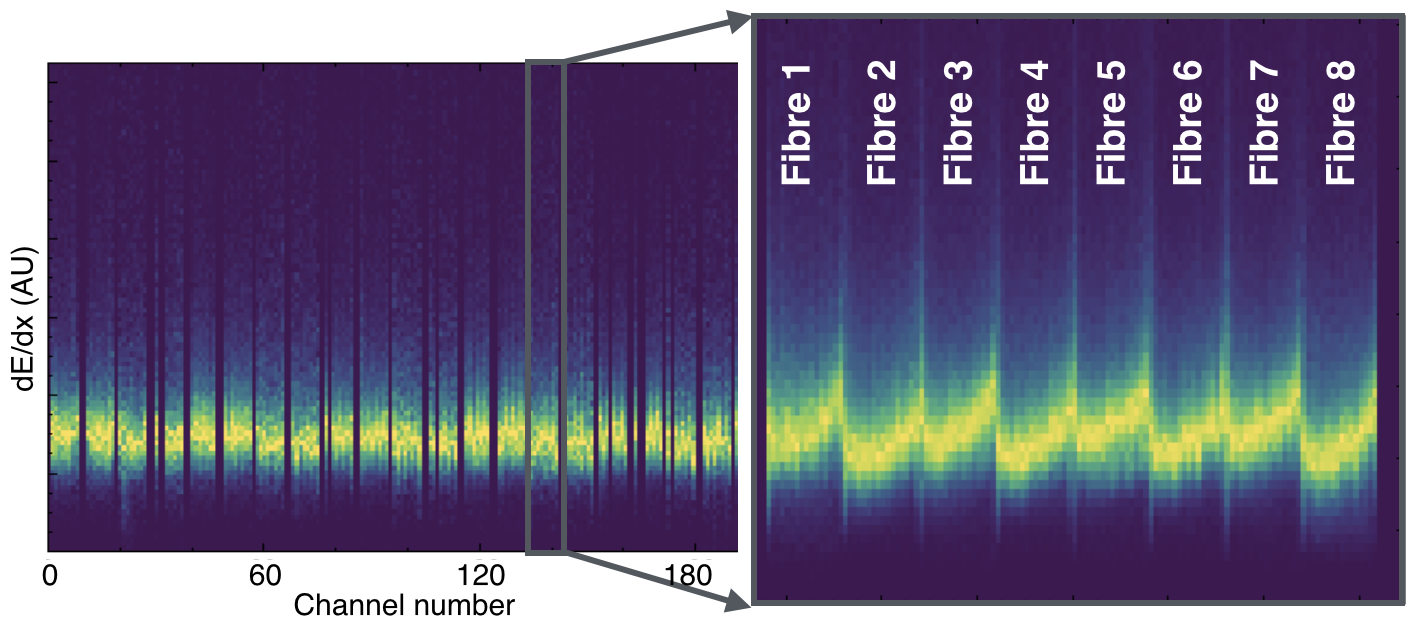}
\caption{Left: $dE/dx$ distribution for the first 200 channels of SM1 (one slice per channel, post channel calibration). Right: zoom on one block of eight channels, showing $dE/dx$ as a function of cube (one slice per cube) along the fibre. The shifts in the distributions are due to attenuation along the fibre.}
\label{fig:attenuation_finder}
\end{center}
\end{figure}

\begin{figure}
\begin{center}
\includegraphics[width=1.0\textwidth]{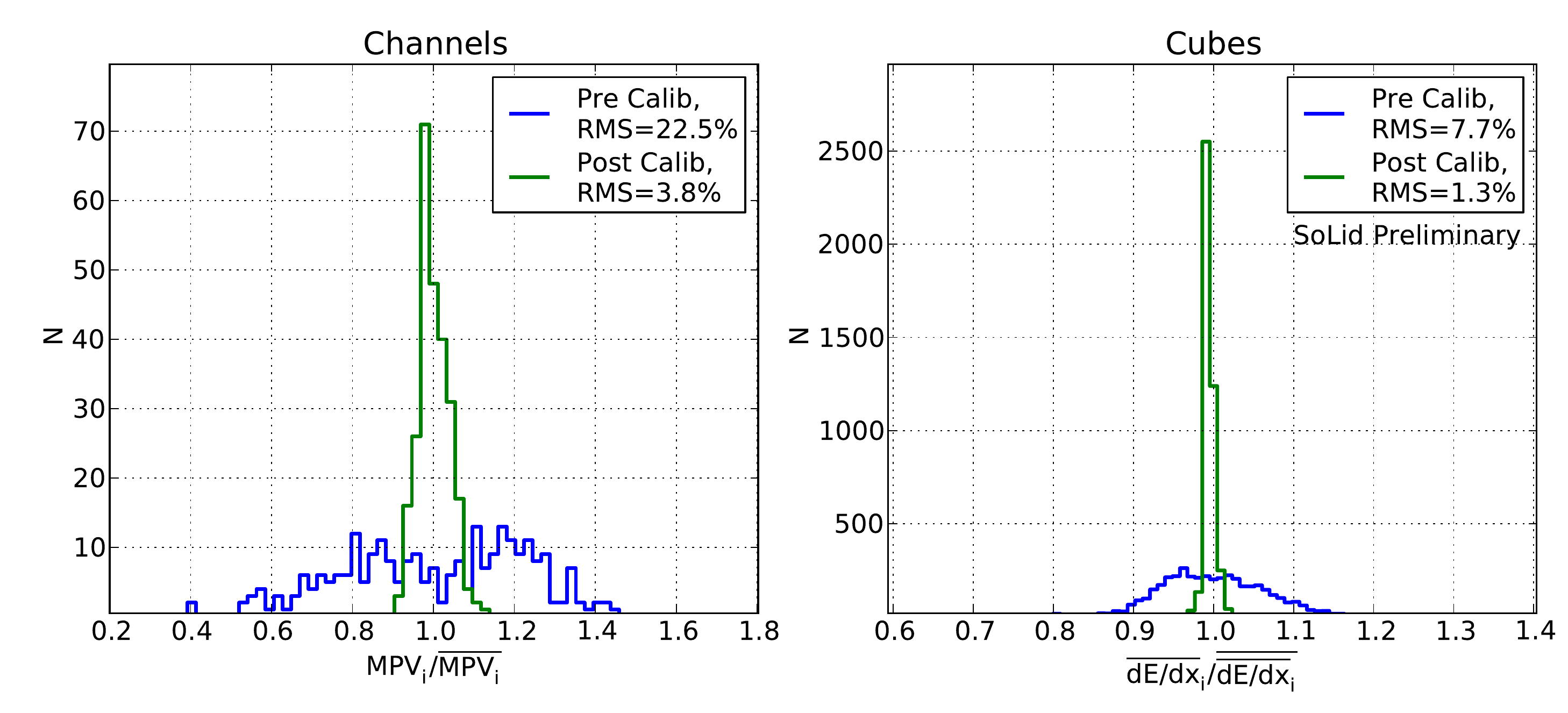}
\caption{Comparison of the spread in channel response (left), and cube response (right), before and after calibrations. In the cube comparison, channel calibration and attenuation corrections have been applied.}
\label{fig:chan_pulls}
\end{center}
\end{figure}

Examples of the $dE/dx$ distributions across the first 200 SM1 channels (post channel calibration) are shown in the left of Figure~\ref{fig:attenuation_finder}. Evidence of the quality of equalisation is shown in the distribution of channel MPVs (normalised to the mean) before and after this offline equalisation, shown in the left of Figure~\ref{fig:chan_pulls}. The spread between channels is reduced from $22.5\%$ to $3.8\%$. The statistical uncertainty of each of the calibration constants is $<1\%$, and it is found that significant Cube-to-Cube variations are the cause of the remaining variations.

\subsection{Fibre Attenuation}
Previous investigations suggest a $\approx20\%$ difference in the variation in light collection efficiency as a function of cube position along a given fibre \cite{celinePre}. This effect is observed when considering the $dE/dx$ distribution for each cube - examples are shown in the right of Figure~\ref{fig:attenuation_finder}, which shows $dE/dx$ as a function of cube, ordered by position along the fibre.

The attenuation can be modelled by the sum of two exponentials (a contribution from light that travels directly down the fibre to the SiPM, and another contribution from the light reflected from the non-instrumented end of the fibre):

\begin{center}
$\frac{\Delta E}{E} (x) = \frac{1}{2}  (e^{-x/a} + re^{-(2x_{fibre} - x)/a})$
\end{center}

\noindent
where $x$ is the position along in the fibre, $\frac{\Delta E}{E}$ is the fraction of light detected, $x_{fibre}$ is the total length of the fibre, $r$ is a reflectivity constant, and $a$ is an attenuation constant. Fitting $r$ and $a$ to data gives:

\begin{center}
$a = 1.31 \pm 0.1$ m$^{-1}$ (stat), and $r = 0.85 \pm 0.03$ (stat)
\end{center}

The average difference (over fibres) between the light correction efficiency between the two ends of the fibre is $24\%$. 

\subsection{Cube-to-Cube Calibration}
A similar procedure to the channel calibration can be performed at the cube level. Each fibre to cube coupling is assigned a scaling calibration constant (therefore two constants per cube), and each constant is floated to equalise the cubes. The distributions of cube $\overline{dE/dx}$ (again, normalised to the mean) for the per cube calibration are shown in the right of Figure~\ref{fig:chan_pulls}. It can be seen that even with channel and attenuation corrections, there is a spread of $8\%$ in cube response. Post cube calibration, this is reduced to $1.4\%$, and this is approaching the statistical uncertainty of the method. 

\subsection{Calibration Summary}
Muons have been used to apply three calibration corrections. For each step, the $dE/dx$ distribution of tracks for each cube has been used. Summarising each step:

\begin{itemize}
\item \textbf{Channel-to-channel calibration}: Offline channel equalisation reduced the spread in channel response from $22.5\%$ to $3.8\%$.This is the most significant correction.
\item \textbf{Attenuation in the fibre}: Light attenuation in the wavelength shifting fibres causes a $24\%$ difference in the light collection efficiency between the ends of the fibre. This is consistent with previous measurements \cite{celinePre}.
\item \textbf{Cube-to-Cube calibrations}: Applying a cube-specific equalisation constant reduced the spread in cube response from from $7.7\%$ to $1.3\%$. This is an acceptable variation given the much larger effect of photon counting statistics on the energy resolution of the cubes.
\end{itemize}

\subsection{Light Yield}
The absolute energy scale is found from comparisons between data and read out simulation \cite{geant4}. This leads to the following light yield:

\begin{center}
1 MeV = 13.0 $\pm$ 0.1 pixel avalanches\footnote{Approximately equal to the number of photons detected.} detected per fibre
\end{center}

In phase 1 of SoLid, the number of fibres per cube will be doubled, giving a light yield of $1 \text{MeV} = 50$ pixel avalanches per cube (using 4 fibres), leading to an enhanced anticipated energy resolution from $20\%$ to $14\%$ at 1 MeV, where the largest contribution is from photon counting statistics. 

\section{Timing Resolution}
\label{sec:timing_resolution}
Muons provide the ability to measure the timing resolution of the detector. Specifically, consider the time residual distribution of track hits with a track: $t_{hit} - t_{track}$. The track time is taken as the average of the hits forming the track, correcting for time-of-flight effects. The track hit time is extracted from the signal waveforms (see \cite{nickPre}), and two methods, with differences in accuracy and computational complexity, have been tested:

\begin{itemize}
\item Timestamp is taken as the bin centre of the peak ADC sample (N.B. the DAQ clock samples at $16$ ns intervals, therefore the expected resolution of this method is $4.6$ ns). 

\item Each signal peak is fitted by a template function from a channel database.
\end{itemize}

\begin{wrapfigure}{R}{0.5\textwidth}
\includegraphics[width=0.5\textwidth]{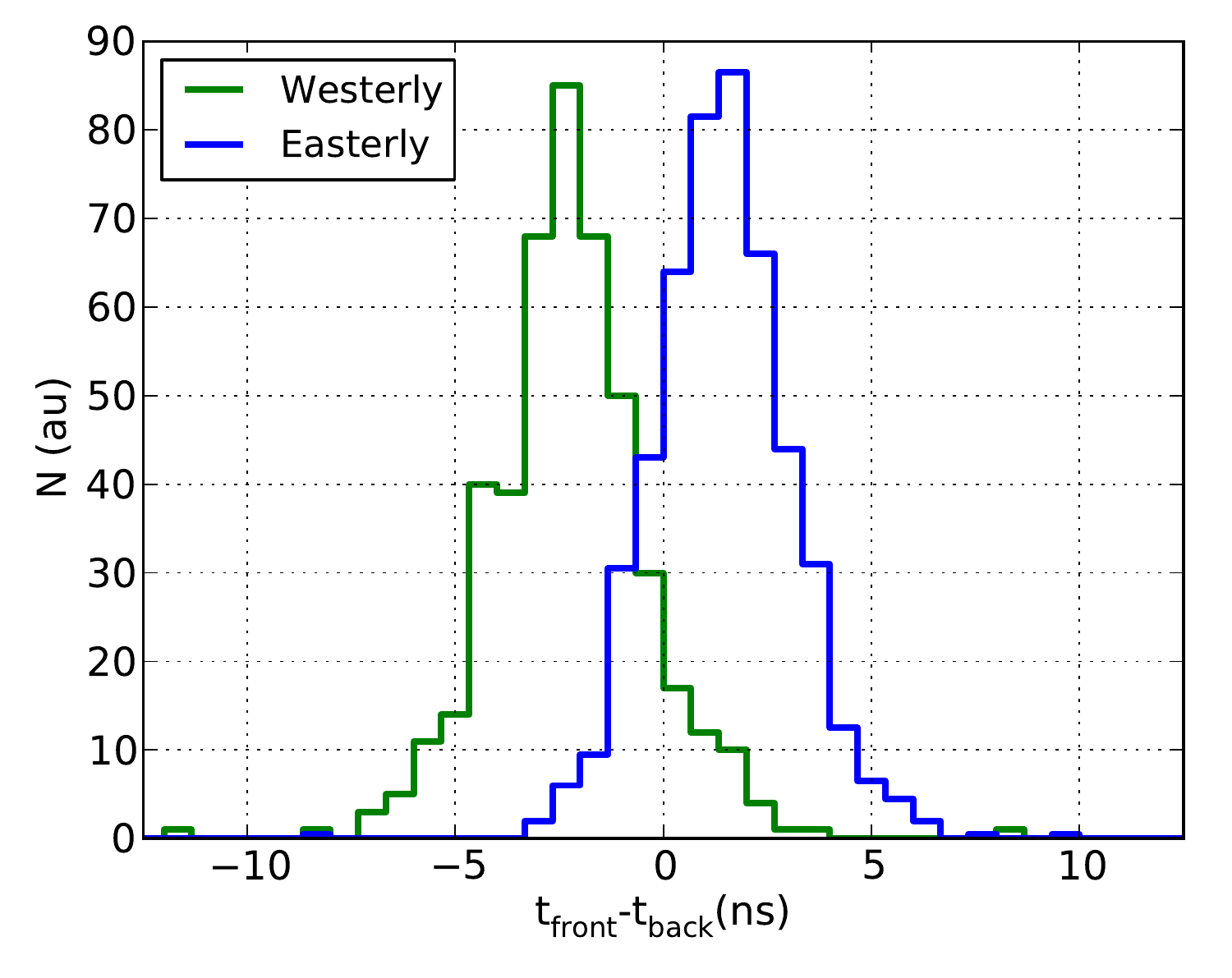}
\caption{Separation between east and west tracks using timing information from the front and back track hits only. Time stamps found by fitting.}
\label{fig:TOF}
\end{wrapfigure}

Taking the time resolution to be the RMS of a Gaussian fit to the residual distributions, the resolutions and computational speeds are summarised in Table~\ref{tab:timeResSummary}. These resolutions imply time of flight effects are noticeable. This can be observed by using the time difference between track hits on the front and back planes of cubes on SM1: $t_{front} - t_{back}$. For tracks entering the detector via the front plane (i.e. from the west), this difference will be negative; for tracks entering the detector via the back plane (i.e. from the east), this time difference will be positive - see Figure~\ref{fig:TOF}. This resolution can be useful for resolving track ambiguities.

\begin{table}
\centering
\begin{tabular}{|p{0.25\columnwidth}|p{0.25\columnwidth}|p{0.4\columnwidth}|}
\hline
\textbf{Algorithm} & \textbf{Resolution $\pm 0.02$ (ns)} & \textbf{Event reconstruction speed decrease} \\
\hline 
Peak bin position  & $5.70$ & $1.0 \times$\\ 
\hline 
Fit with templates & $1.53$ & $6.0 \times$ \\ 
\hline 
\end{tabular} 
\caption{Summary of time resolutions.}
\label{tab:timeResSummary}
\end{table}

\subsection{Conclusions}
The most recent SoLid prototype SM1 has been successfully commissioned in Spring 2015, and has recorded a rich data set including reactor on/off comparisons, as well as radioactive source data for both electromagnetic and neutron calibration. Using the fine segmentation of the detector, muons can be reconstructed and used to calibrate the electromagnetic response of the detector to the cube level, giving a spread of $<2\%$ in cube response post calibration. This is significantly below the level of the anticipated energy resolution of the detector. The absolute energy scale was found to be 26 pixel avalanches per MeV absorbed in PVT. Muons also provide a means to measure the timing response of the detector, with per-hit timing resolution $\sigma_{Time} = 1.5 \pm 0.2$ ns.

\end{document}